\documentclass[doublespacing]{elsart}
\usepackage{graphicx,amssymb}
\journal{New Astronomy}
\def\astrobj#1{#1}
\begin{document}

\begin{frontmatter}
\title{Photometric Study of Neglected Eclipsing Binary \astrobj{GSC 3576-0170}}

\author[]{Liyun Zhang$^{1,2}$}
\ead {liy\_zhang@hotmail.com, {\it phone:} +86 851 362 7662, {\it
fax:} +86 851 362 7662}
\author[]{Jianghong Jing$^{1,2}$}
\author[]{Yanke Tang$^{3}$}
\&\author[]{Xiliang Zhang$^{4}$}
\address{$^{1}$College of Science, Department of Physics, Guizhou University,
Guiyang 550025, PR China}

\address{$^{2}$NAOC-GZU-Sponsored Center for Astronomy Research, Guizhou University,
Guiyang 550025, PR China}

\address{$^{3}$Department of Physics, Dezhou University, Dezhou 253023,
  PR China}

\address{$^{4}$National
Astronomical Observatories, Yunnan Observatory, Chinese Academy of
Sciences, Kunming, 650011, PR China}

\begin{abstract}
The new multi-color $BVRI$ photometric light curves of the
short-period eclipsing binary \astrobj{GSC 3576-0170} were obtained
on two consecutive nights (October 5 and 6, 2009). With the 2003
version of Wilson-Devinney program, the precise photometric
solutions are derived for the first time. The result shows that
\astrobj{GSC 3576-0170} is a semi-detached binary system with a
large temperature difference of approximately 1490~K. The
light-curve distortions are further explained by a hot spot on the
secondary component through mass transfer via a stream hitting the
facing surface of the secondary component. By analyzing all
available light minimum times, we also derived an update ephemeris
and found for the first time a possible periodic oscillation with an
amplitude of 0.0038 days and a period of 4.3 years. The periodic
oscillation could be explained either by the light-time effect due
to a presumed third component or by magnetic activity cycle of the
system.
\end{abstract}
\end{frontmatter}
\section{Introduction}
\label{intro} \astrobj{GSC 3576-0170}(P$_{orb}$=$0^{d}$.405, G1 V)
is a new near-contact solar-type eclipsing binary, which displays
asymmetric light curves (two different light
maxima) (Nelson et al. 2006). Therefore, it is a very intriguing target for understanding the property of the system.\\
\indent \astrobj{GSC 3576-0170} was discovered to be variable by
Nelson et al. (2006). They attempted a preliminary light-curve
synthesis with Wilson-Devinney modeling and preferred a detached
model with a mass ratio in the range $0.15-0.35$ and an orbital
inclination of 65-$70^{\circ}$. For the period variation, they
invoked a quadratic function to fit the values of the observational
times of light minimum - calculational times of light minimum
$(O-C)$. Subsequently, new minimum times of GSC 3576-0170 were
published by several astronomers (Nelson et al.
2006; H$\ddot{u}$bscher \& Walter 2007; H$\ddot{u}$bscher et al. 2006; Br\'{a}t 2007; etc.).\\
\indent In this paper, we present our new $B, V, R,$ and $I$ LCs of
\astrobj{GSC 3576-0170} which were analyzed using the 2003 version
of Wilson-Devinney code (Wilson \& Devinney 1971; Wilson 1979, 1990,
1994; Wilson
\& Van Hamme 2004). We accumulated all available times of light minimum and discussed the period change.\\
\begin{table}
\caption{Minimum times of \astrobj{GSC 3576-0170}.} \tabcolsep
0.30truecm
\renewcommand\arraystretch{0.6}
\begin{tabular}{lrrc}
\hline \hline \multicolumn{1}{l}{JD(Hel.)} &
\multicolumn{1}{c}{Cycle} & \multicolumn{1}{c}{(O-C)}  &
\multicolumn{1}{c}{References} \\
\hline
2452794.863     & $-2.5$ & $-0.0001$ & 1\\
2452795.8716     & $0.0$ &$-0.0038$  & 1\\
2452799.9230     & $10.0$ & $-0.0024$  & 1 \\
2452802.5542       &$16.5$ &$-0.0039$  &1,4\\
%2452802.5543&$16.5$&$-0.0037$&4\\
2452806.8076     &$27.0$ &$-0.0029$   &1\\
2452807.8210&$29.5$&$-0.0020$& 1\\
2452812.4781&$41.0$&$-0.0026$& 1,4\\
%2452812.4781&$41.0$&$-0.0025$&4\\
2452826.8600&$76.5$&$0.0017$&1\\
2452829.4887&$83.0$&$-0.0021$&4\\
2452831.5151&$88.0$&$-0.0007$&4\\
2452863.5105&$167.0$&$-0.0007$&4\\
2452864.5304&169.5&0.0067&4\\
2452867.5607&177.0&$-0.0006$&4\\
2452868.3701&179.0&$-0.0012$&4\\
2452946.3385&371.5&$0.0037$&4\\
2453215.4640&1036.0&0.0034&4\\
2453216.4767&1038.5&0.0036&4\\
2453217.4888&1041.0&$0.0032$&4\\
2453221.5370&1051.0&0.0013&4\\
2453263.8659&$1155.5$&0.0072&1 \\
2453264.6735&$1157.5$&$0.0048$&1\\
2453305.7787 &$1259.0$&0.0020&1\\
2453612.3645&2016.0&$-0.0011$&4\\
2453612.5733&2016.5&0.0052&4\\
2453621.4824&2038.0&0.0042&4\\
2453837.9506&$2573.0$&$-0.0028$&1\\
2453852.9370&$2610.0$&$-0.0016$&1\\
2453900.7278&$2728.0$&$-0.0014$&1\\
2453941.8337&$2829.0$&$-0.0035$&1\\
2453943.8605&$2834.5$&$-0.0017$&1\\
2454073.2580&$3154.0$&$-0.0033$&2\\
2454270.4947&$3641.0$&$-0.0041$&3\\
2455109.0636&5711.5&$0.0019$&5\\
2455110.0739&5714.0&$-0.0004$&5\\
%\hline
%2451016.3973&$-766.5$&0.0031& Ekmek\c{c}i \& Ak., (2001)\\
\hline
\end{tabular}
\small
\renewcommand\arraystretch{0.1}
\\References - (1) Nelson et al. 2006; (2) H$\ddot{u}$bscher \& Walter
2007; (3) Br\'{a}t 2007; (4) H$\ddot{u}$bscher et al. 2006; (5)
Present paper.
\end{table}

\section{New CCD photometric observations of \astrobj{GSC 3576-0170}}
New CCD observations were made on October 4 and 5, 2009 with the 85
cm telescope at Xinglong Station of National Astronomical
Observatories of China (NAOC). The telescope was equipped with a
standard Johnson-Cousin-Bessel multi-color CCD photometric system
built on the primary focus (Zhou et al. 2009). The camera was
equipped with a 1024 $\times$ 1024 CCD and the field of view was
about $16^{\prime}.5\times 16^{\prime}.5$ (Yang et al. 2009). All
CCD images were reduced by means of IRAF package in the standard
fashion. The star ($\alpha_{2000}=20:23:44.56$;
$\delta_{2000}=46:51:33.9$) and \astrobj{GSC 3576-0970}
($\alpha_{2000}=20:24:04.04$; $\delta_{2000}=46:53:21.8$) were
chosen near the target as comparison and check stars, respectively.
The magnitude of these stars were determined using the Daophot
subpackage of IRAF and the errors were found to be lower than 0.01
mag in all bands. Corresponding light curves are displayed in figure
1, where the phases of data points are
calculated using the new light ephemeris (the equation~1).\\
\begin{figure}
  \begin{center}
    \includegraphics*[width=25pc, height=20pc]{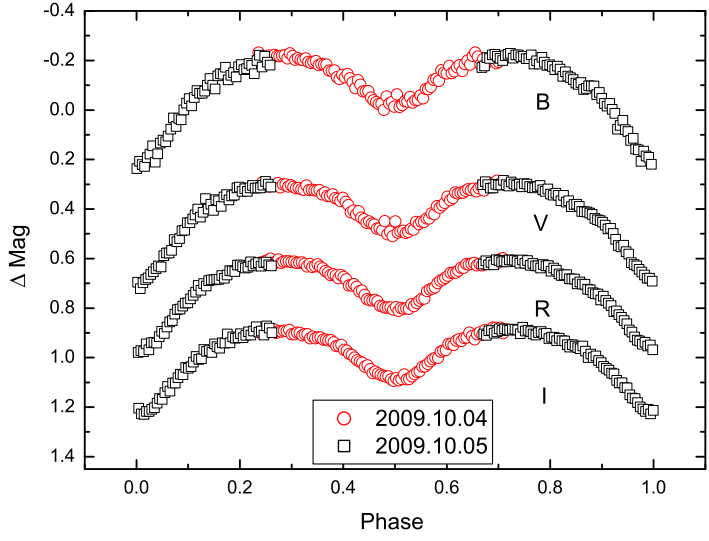}
   \end{center}
    \caption{CCD photometric observations of \astrobj{GSC 3576-0170}.}\label{fig:1}
\end{figure}

\section{Orbital Period Variation for \astrobj{GSC 3576-0170}}
\indent From our observations, a new secondary minimum
(2455109.0636$\pm$0.0025) and a primary minimum (2455110.0739$\pm$
0.0015) for \astrobj{GSC 3576-0170} are obtained by means of
parabola fitting. To discuss periodic variation, we compiled all
available light minima, which are listed in table 1. Using the
linear least-squares method, an update linear ephemeris is given by
equation(1):
\begin{eqnarray}
Min.I=HJD2452795.8754(\pm 0.0007)+0^{d}.4050051(\pm 0.0000004)E
\end{eqnarray}
Based on our new ephemeris, the $O-C$ values of minimum times are
also listed in table~1. The calculation and the plot are provided in
figure~2. As can be seen from figure 2, a cyclic variation might be
present. To clarify the period change, we used a sinusoidal function
for fitting which led to the following equation:
%\begin{eqnarray}
%Min.I~=~JD(Hel.)2451671.0427(\pm 0.0003)
%\nonumber\\
%+0^{d}.405012(\pm 0.000001)E \nonumber -2^{d}.3(\pm
%0.4)\times10^{-9}\times E^{2}.
%\end{eqnarray}
\begin{eqnarray}
%Min.I~=~HJD2452795.8754(\pm 0.0007)
%+0^{d}.4050051(\pm 0.0000004)E\\
O-C=0^{d}.0038(\pm 0.0006)\times\sin(0^{\circ}.093(\pm
0^{\circ}.005)\times E-27^{\circ}.1(\pm 1^{\circ}.3)).
\end{eqnarray}
The fit is also plotted in figure~2 with solid line. The sinusoidal
term of equation (2) reveals a periodic oscillation with an
amplitude of $A=0^{d}.0038$. Using
T=$360^{\circ}$$\times$$P$/$\omega$, where $P$ denotes the orbital
period of \astrobj{GSC 3576-0170} in years and $\omega$ is the
coefficient of E; the period of the oscillation T is determined to
be $4.3\pm0.2$ years. The size of the periodic variation $\Delta
P$/P=1.5$\times$$10^{-5}$ can be computed in days using the equation
$\Delta P/P=A\times\sqrt{2[1-cos(2\pi\times P/T)]}$/P
(Rovithis-Livaniou et al. 2000), where A is the semi-amplitude of
the light-time effect, and T is the period of the
light-time effect. \\
%This ephemeris suggests a continuous secular decrease of the period at a rate of dp/dt=$-2.5\times 10^{-8}dyr^{-1}$\\

\section{Photometric analysis}
For our light curves of \astrobj{GSC 3576-0170}, two light maxima
are almost equal in $BVRI$ bands (the differences are all less than
0.01 mag). This implies that it can be used to determine reliable
photometric parameters. During the solution process, the four light
curves are simultaneously analyzed. The preliminary values of these
orbital parameters are taken from the preliminary photometric
solutions derived by Nelson et al. (2006). The details of the
procedure of photometric solution are similar to those of our
previous work of \astrobj{RT And} (Zhang \& Gu 2007), \astrobj{DV
Psc} (Zhang et al. 2010),
and \astrobj{KQ Gem} (zhang 2010).\\
%Since the eclipses were obviously partial, it is not
%possible to determine precisely mass ratio based on photometric data
%alone (Terrell \& Wilson 2005).
\indent To be on the safer side, we explored a wide range based on
the estimation of Nelson et al. (2006). The mass ratios were fixed
at 0.2, 0.25, 0.3, 0.35, 0.4, 0.45, 0.5, 0.6, and 0.7, respectively.
For each assumed value of q. the light curves were solved based on
mode 2~(detached mode) of Wilson and Devinney's DC Program.
According to the spectral type G1 V of the primary (Nelson et al.
2006) and the spectra-effective temperature relation of Cox (2000),
the effective temperature of the primary $T_{1}$ was fixed at 5865.
During the solution process, we assumed synchronous rotation and
zero eccentricity. Simple treatment was used to compute the reflect
effect, and the linear limb-darkening law was employed to compute
the limb-darkening effect. The bolometric albedo $A_{1}$=$A_{2}$=0.5
(Rucindki 1973), the limb-darkening coefficients $x_{1B}$~=~0.742
$x_{2B}$~=~0.906, $x_{1V}$~=~0.608 $x_{2V}$~=~0.763,$x_{1R}$~=~0.503
$x_{2R}$~=~0.629, $x_{1I}$~=~0.413 $x_{2I}$~=~0.512 (Van Hamme
1993), and the gravity-darkening coefficients
$g_{1}$~=~$g_{2}$~=~0.32 (Lucy 1967) are set for the primary and the
secondary in the usual manner. The solutions from several runs were
obtained for all assumed mass ratios. The relation of the sum of
weighted square deviation ($\sum_{i}(O-C)_{i}^{2}$) with mass ratio
q is illustrated in figure 3, where the lowest value of $\sum$ was
found at q = 0.45. Therefore, it indicates that the most likely mass
ratio appears to be approximately 0.45. The photometric solutions
for 0.45 are listed in table 2 and corresponding light curves are
plotted with dashed lines in
figure 4.\\
% After many runs, the solutions for several
%assumed value of mass ratio q are obtained for \astrobj{GSC 3576-0170}. For
%each q, the calculation stars at mode 2 (the detached mode), the
%sums of weighted square deviation ($\sum_{i}(O-C)_{i}^{2}$) for all
%the assumed values of q are shown in Figure 3. The minimum $\Sigma$
%is achieved at q~=~0.45, Therefore we perform a differential
%correction so that it converges by making q an adjustable parameter
%and by
%choosing q~ = ~0.45 as the initial values for GSC 3576 -0170. \\
%\indent When performing the differential correction calculation, the
%adjustable orbital parameters are the orbital inclination $i$, the
%temperature of the secondary $T_{2}$, the dimensionless potentials
%of the two components $\Omega _{1}$ and $\Omega _{2}$, and the
%monochromatic luminosity of the primary $L_{1}$ deriving from the
%approximate Kurucz atmosphere model option of the Wilson-Devinney
%program (Kurucz 1993). The preliminary values of these orbital
%parameters are taken from photometric solution derived by Zhang \&
%Zhang (2007). The details of the procedure of photometric solution
%is similar to that of photometric solution of RT And (Zhang \& Gu
%2007) DV Psc (Zhang et al., 2010) and KQ Gem (zhang 2010;).\\
\indent Comparing the observed LCs with the theoretical ones for the
likely mass ratio(see figure~4) shows enhancement (brighter) on the
shoulders of the secondary eclipse. This kind of enhancement may be
caused by mass transfer from the primary component. To account for
this phenomenon, we adopted a semi-detached model with the addition
of a hot spot on the secondary. For this purpose, we changed the
mode from 2 to 4 (semi-detached mode with lobe-filling primary), and
continued the DC calculation until the final photometric solution of
was derived. At the same time, the mass ratio q were released as a
free parameter and continue the DC calculation based on our initial
solutions of mode 2. To avoid the correlations among the adjusted
parameters, they were divided into two subsets: the orbital
parameters and spot parameters. For the spot parameters, the spot
latitude is assumed to be $90^{\circ}$, which means that spot center
is on the equator of the component. The spot temperature is assumed
to be the same as that of the primary (5860~K) due to the mass
transfer, that is transferred from the primary component, would be
about that of the primary component. Therefore, we only adjusted the
remaining two
parameters of the spot, the spot longitude $longitude _{spot}$ and the radius $radius _{spot}$. % The preliminary longitudes of the
%spots are determined from the centers of the depression of LCs.
%While, the radius of the spot is assumed by the theoretical LCs
%fitting the observed ones, especially in the phase ranges of the
%photometric distortions.
The solutions from several runs were derived. Corresponding
photometric solutions are listed in table 2. The theoretical light
curves are plotted in figure 4 with solid lines, and corresponding
configurations of \astrobj{GSC 3576-0170} in phases
0.0, 0.25, 0.5 and 0.75 are shown in figure 5. \\
%\subsection{Analysis of the light curve variation}

\begin{table}
\tabcolsep 0.36truecm
\renewcommand\arraystretch{1.0}\caption{The results of LC analysis for
\astrobj{GSC 3576-0170}.}
\begin{tabular}{lcccc}
\hline \hline \multicolumn{1}{l}{Element} &
\multicolumn{1}{c}{detached}& \multicolumn{1}{c}{semi-detached}
%& \multicolumn{1}{c}{detached}%&
\\
\hline
$T_{1}$&5865K a&5865K a\\
q& 0.450a& 0.467$\pm$0.003 \\
i & $64.^{\circ}32$$\pm$0.69 & $64.^{\circ}63$$\pm$0.14 \\
\hline
$T_{2}$& 4369$\pm$46~K  &   4375$\pm$27   \\
$\Omega_{1}$ & 2.702$\pm$0.004 &2.798a      \\
$\Omega_{2}$ & 2.880$\pm$0.023  &2.790$\pm$0.005     \\
$L_{1B}/(L_{1}+L_{2})_{B}$  & 0.9587$\pm$0.0002 &  0.9444$\pm$0.0002   \\
$L_{1V}/(L_{1}+L_{2})_{V}$ & 0.9302$\pm$0.0004 & 0.9073$\pm$0.0004   \\
$L_{1R}/(L_{1}+L_{2})_{R}$  & 0.9088$\pm$0.0005 &   0.8807$\pm$0.0005           \\
$L_{1I}/(L_{1}+L_{2})_{I}$  & 0.8830$\pm$0.0007 &   0.8485$\pm$0.0006     \\
$r_{1}$(pole)& 0.4368$\pm$0.0008 &0.4200$\pm$0.0005   \\
%$r_{1}$(point)& 0.3457$\pm$0.0006 &  0.3434$\pm$0.0006 &  0.3446$\pm$0.0006 &0.3447$\pm$0.0006        \\
$r_{1}$(side)& 0.4684$\pm$0.0010 &  0.4466$\pm$0.0005 \\
$r_{1}$(back) & 0.5015$\pm$0.0014 &    0.4743$\pm$0.0005  \\
$r_{2}$(pole)  & 0.2743$\pm$0.0037 &  0.2984$\pm$0.0013    \\
%$r_{2}$(point) & 0.2281$\pm$0.0003 &   0.2274$\pm$0.0003 &  0.2281$\pm$0.0003 & 0.2293$\pm$0.0003 \\
$r_{2}$(side)  &0.2840$\pm$0.0043   & 0.3119$\pm$0.0016    \\
$r_{2}$(back)     & 0.3073$\pm$0.0060 &0.3470$\pm$0.0027   \\
$Latitude_{spot}$     & -   &$90^{\circ}$ a  \\
$Longitude_{spot}$     & - &   $181.^{\circ}9$$\pm$$0.^{\circ}2$   \\
$Radius_{spot}$     & - &  $27.^{\circ}8$$\pm$$0.^{\circ}5$   \\
$Temperature_{spot}$     & - &    5865~K a       \\
%$Latitude_{spot1}$     & $90^{\circ}$ a &$90^{\circ}$ a    \\
%$Longitude_{spot2}$& $202.4^{\circ}$$\pm$$0.1^{\circ}$ &   $190.5^{\circ}$$\pm$$2.6^{\circ}$    \\
%$Radius_{spot2}$& $21.2^{\circ}$$\pm$$0.4^{\circ}$ &  $24.8^{\circ}$$\pm$$4.5^{\circ}$     \\
%$Temperature_{spot1}$     & 6500~K a &    6500~K a       \\
\hline
$\Sigma(O-C)_{i}^{2}$ & 0.7188&0.2694 \\
\hline
\end{tabular}\\
Parameters not adjusted in the solution are denoted by a mark ``a".
\end{table}
\begin{figure}
  \begin{center}
    \includegraphics*[width=25pc, height=20pc]{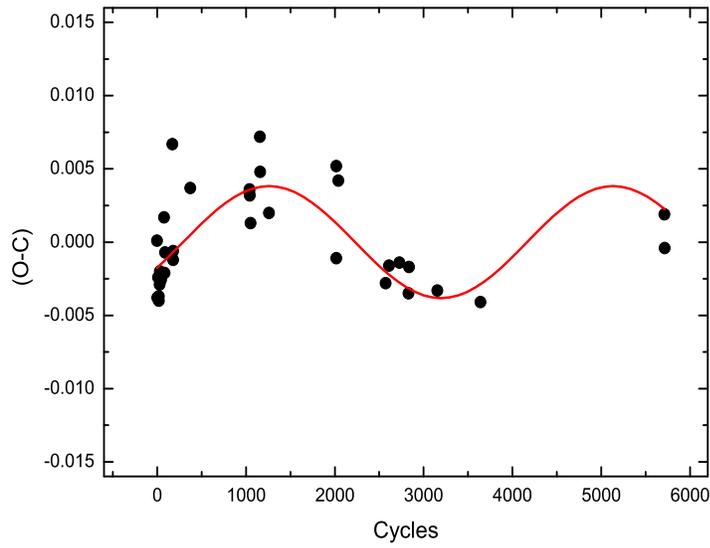}
   \end{center}
    \caption{The O-C diagram for the minimun
times of \astrobj{GSC 3576-0170}. The solid line represents the sin
fitting.}\label{fig:2}
\end{figure}
\begin{figure}
  \begin{center}
    \includegraphics*[width=25pc, height=20pc]{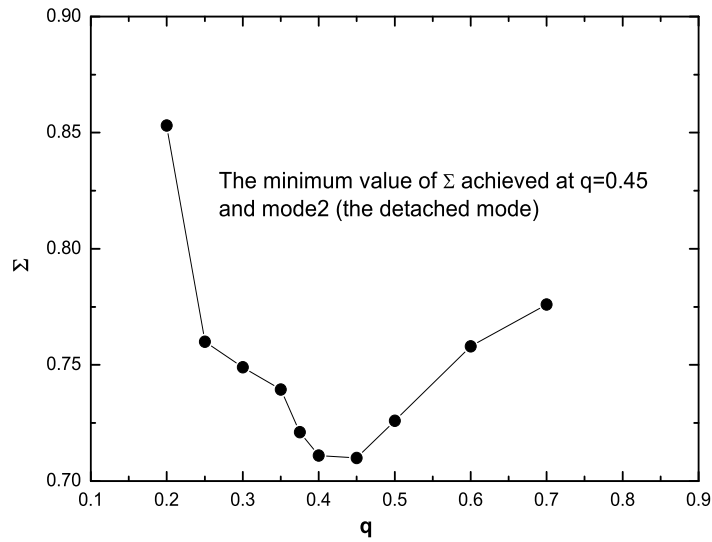}
   \end{center}
    \caption{$\sum$ -- q curves of \astrobj{GSC 3576-0170}}\label{fig:1}
\end{figure}
\begin{figure}
  \begin{center}
    \includegraphics*[width=25pc, height=20pc]{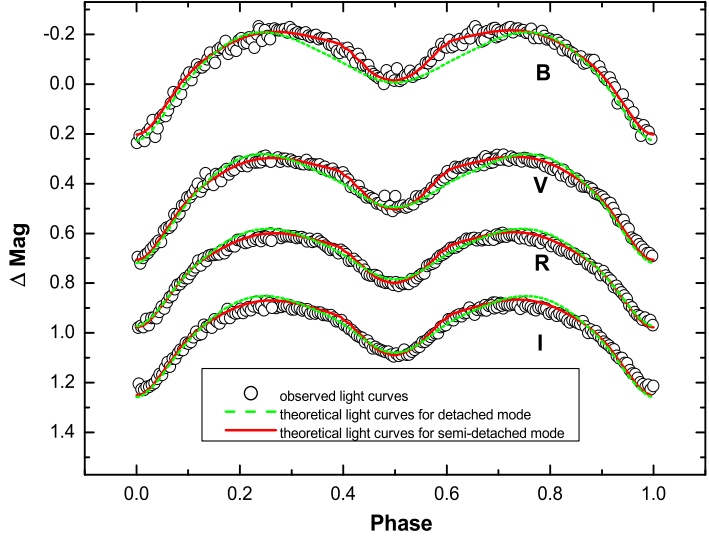}
   \end{center}
    \caption{The observational and theoretical light curves of \astrobj{GSC 3576-0170}.
    The circles represent the observational data. The dashed and solid lines
    represent the theoretical light curves for detached and semi-detached mode, respectively.}\label{fig:1}
\end{figure}
\begin{figure}
  \begin{center}
    \includegraphics*[width=25pc, height=18pc]{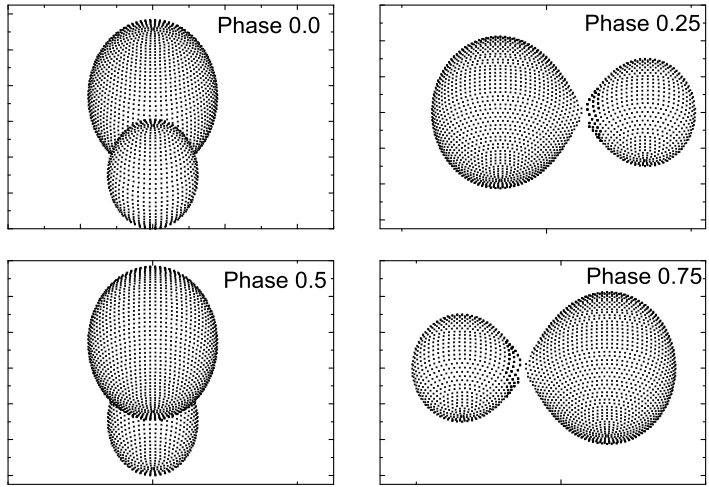}
   \end{center}
    \caption{The configurations of \astrobj{GSC 3576-0170} in phases 0.0, 0.25, 0.5
and 0.75.}\label{fig:1}
\end{figure}

\section{Discussions and Conclusions}
Our new LCs and the period variation of \astrobj{GSC 3576-0170} are
included in present investigation.
\subsection{Cyclic magnetic activity or the light-time effect
due to the third Body?} For \astrobj{GSC 3576-0170}, there seems to
be the sinusoidal oscillation of the O-C residuals. The oscillating
characteristic may be caused by the light-time effect due to the
existence of the third body orbiting the eclipsing binary or a possible result of magnetic activity cycles of the system.\\
 \indent On one hand, the period oscillation may be explained by the light-time effect. Assuming that the orbital of the presumed third body is
circular, we can obtain the mass function for the third body
f($M_{3}$)=0.016($\pm$0.001) M$\odot$ using the following equation:
\begin{equation}
f(m)
=\frac{4\pi^{2}}{G{T}^{2}}\times(a\prime_{12}\sin{i^{\prime}})^{3}=\frac{(M_{3}\sin{i_3^{\prime}})^{3}}
{(M_{1}+M_{2}+M_{3})^{2}}.
\end{equation}
%The mass function of the third body can be computed to f(m)=00.
%
%\begin{equation}
%f(m_3)=\frac{(M_{3}\sin{i_3^{\prime}})^{3}}
%{(M_{1}+M_{2}+M_{3})^{2}}
%\end{equation}
where $M_{1}$, $M_{2}$ and $M_{3}$ are the masses of the binary, and
the third body, respectively. The mass of the third component can
also be calculated with the above equation, which depends on the
orbital inclination. The minimal mass $M_{3,min}$ is 0.38 M$\odot$
when $i^{\prime}$=$90^{\circ}$. \\%Because the third mass is much
%smaller than the mass of the more massive component (the primary),
%its light contribute little to the total luminosity and
%can not be detected now. This suggests that a tertiary component may exist.\\
\indent
%\begin{equation}
%B^{2}\simeq\frac{1}{6} \frac{GM^{2}}{R^{4}} (\frac{a}{R})^{2}
%\frac{\Delta p}{p}
%\end{equation}
%1.024 M$\odot$; the radius is 1.06 R$\odot$ a=2.37 R$\odot$ P=0.405
%p3=4.3 $\frac{\Delta p}{p}$=1.5*$10^{4}$;
On the other hand, the period oscillation may be accounted by
magnetic cycle of the system (Applegate 1992; Lanza et al. 1998;
Lanza \& Rodon$\grave{o}$ 1999). The orbital period change
corresponding to a variation of the quadrupole moment is given by
$\Delta Q$~=~-($\Delta p$/p)$\times$($Ma^{2}$/9) (Applegate 1992)
where a is the semi-major axis of the binary orbit and M is the mass
of the active component. Since no spectroscopic solutions are
available in literature \astrobj{GSC 3576-0170}, its absolute
parameters can not be directly determined. We estimated the primary
mass 1.024 as M$\odot$, and the radius as 1.06 R$\odot$ by assuming
the primary component to be a normal and main-sequence GIV star.
Based on our photometric solutions, the mass of the secondary
component is $M_{2}$ = 0.478 M$\odot$ and the separation between the
two components is a~=~2.37 R$\odot$. Therefore, the quadrupole
moment $\Delta
Q$ is calculated to be 0.5$\times$$10^{50}$ g $cm ^{2}$ for the primary.%The
%sinusoidal variation may result from cyclic magnetic activity.
%Strong magnetic activities on the surface of one or both stars may
%produce this kind of variation. The required variation of the
%quadrupole moment can be estimated to be 4.6 $10^{50}$ g $cm^{2}$
%for the primary components, respectively. It similar to the
 This value is similar to the typical values for active binaries
(Lanza \& Rodono 1999). Therefore, the magnetic
activity cycle is a possible mechanism to explain the variation of period.\\
\subsection{Long-term change and their evolutionary status}
For our observations, the two light maxima are basically equal.
However, the light curves show that the first quadrature is brighter
than the second one in 2003 (Nelson et al. 2006). These indicate a
variability on a time scale of about six years. This variation might
be attributed to surface magnetic activity of one or both stars.
More importantly, GSC 3576-0170 is perhaps associated with possibly
thick deep convection according to their spectral types G1 V. Thus,
it confirms that magnetic activities may produce the
sinusoidal variation of the period.\\
%\indent From our results, a hot-spot model being on the secondary
%star is successful in representing the distortions of the light
%curves of \astrobj{GSC 3576-0170}. For the orbital parameters, the
%orbital inclination i=$64.^{\circ}63$$\pm$0.14 and the mass ratio
%0.467, the dimensionless potential of the secondary components
%$\Omega _{2}$ is 2.790. The contribution of the primary component to
%the total light is 0.9444 in $B$, 0.9073 in $V$, 0.8807 in $R$ and
%0.8485 in $I$ band. Since starspots are the sites of mass transfer,
%the total percent area covered by hot spot can provide an indirect
%measure of mass transfer. For \astrobj{GSC 3576-0170}, the hot spot
%covers about 5.8($\pm$0.3)\% of the surface on
%the secondary component.\\

%the radius is 1.06 R$\odot$
% \indent A cycle was determined from the analysis of the orbital
% period. The variation period coincides with 52.2 given by Pri.
% as the times of light minimum in the latest five years were added,
% a cyclical variation with a period of was confirmed. The light time
% effect is possible reason for the cyclical variation. The
% quasi-sinusoidal O-C curve allow us to assume that the orbital of
% the tertiary component is circular. The mass of tertiary component
% m3 depends on the orbital inclination. m3 has minimum values when
% i=$90^{\circ}$. It light contribute little to the total luminosity and cannot
% be detected. This suggests that a tertiary component may exist. \\
%\indent %With all available times of light minimum, a new ephemeris
%was derived and the O-C residuals of \astrobj{GSC 3576-0170} were
%analyzed.
\indent Our result showed that \astrobj{GSC 3576-0170} is a
marginal-contact binary system with a large temperature difference
of about 1490~K between the two components, which belongs to the
subclass of V1010 Oph binaries (Shaw 1994). It is similar to several
semi-detached systems with a possible third body, such as
\astrobj{GW Tau} (Zhu \& Qian 2006). They may be potential
candidates of binaries that will become overcontacted or remain in
the broken-contact in accordance with the thermal relaxation
oscillation theory of overcontact binaries (Lucy 1976; Flannery
1976; Robertson \& Eggleton 1977; etc.).\\
Of course, our solutions are based on photometric observations only.
For better understanding of the properties and the evolutionary
state of \astrobj{GSC 3576-0170}, spectroscopic observations are
needed. Moreover, the O-C residual of \astrobj{GSC 3576-0170} is
about seven years and there is an observational gap. Therefore, it
is too early to decide the character of the periodic variation, and
it might take another 10-20 years are needed for confirmation.\\
{\bf Acknowledgements}
 The authors gratefully acknowledge the assistance provided for the 85 cm telescope at Xinglong station. We also thank Profs. X., Zhou, A. Y. Zhou, X.J. Jiang,
and Y. H. Zhao, for allocation of time for observation and their
kind helps during the visit to NAOC. This work was partly supported
by GuiZhou University under Grant No. 2008036, GuiZhou Natural
Science Foundation 20092263, Shandong Natural Science Foundation
(ZR2009AM021) and the Joint Fund of Astronomy of the National
Natural Science Foundation of China and the Chinese Academy of
Sciences Grant No. 10978010, Natural and Scientific funding
supported by Department of Education of Guizhou Province No.
20090130, and Dezhou University Foundation (402811).

\end{document}